\documentclass[aps,prd,preprintnumbers,twocolumn]{revtex4}
\usepackage{epsfig, epsf}
\def\lsim{\raise0.3ex\hbox{$<$\kern-0.75em\raise-1.1ex\hbox{$\sim$}}}
\def\gsim{\raise0.3ex\hbox{$>$\kern-0.75em\raise-1.1ex\hbox{$\sim$}}}
\def\be{\begin{equation}}
\def\ee{\end{equation}}
\def\ba{\begin{eqnarray}}
\def\ea{\end{eqnarray}}
\def\bea{\begin{eqnarray}}
\def\eea{\end{eqnarray}}
 2
\def\m0{m_{D0}}
\def\bx{{\bf x}}
\def\b0{{\bf 0}}

\newcommand{\tr}{{\rm Tr}}

\begin{document}
\preprint{BI-TP 2004/16}
\preprint{~}

\title{Static quark anti-quark free energy and the running coupling at finite
temperature}
\author{O. Kaczmarek, F. Karsch}
\affiliation{
Fakult\"at f\"ur Physik, Universit\"at Bielefeld,\\
\qquad  P.O. Box 100131, D-33501 Bielefeld, Germany\\
}
\author{P. Petreczky}\thanks{Goldhaber Fellow}
\affiliation{Physics Department, Brookhaven National Laboratory, Upton, NY 11973 USA}
\author{F. Zantow}
\affiliation{
Fakult\"at f\"ur Physik, Universit\"at Bielefeld,\\
\qquad  P.O. Box 100131, D-33501 Bielefeld, Germany\\
}
\date{\today}
\begin{abstract}
We analyze the free energy of a static quark anti-quark pair
in quenched QCD at short and large distances. From this we deduce 
running couplings, $g^2(r,T)$, and determine the length scale that separates 
at high temperature the short distance perturbative regime from the large 
distance non-perturbative regime in the QCD plasma phase. 
Ambiguities in the definition of a coupling beyond the perturbative 
regime are discussed in their relation to phenomenological
considerations on heavy quark bound states in the quark gluon plasma.
Our analysis suggests that it is more appropriate to characterize the
non-perturbative properties of the QCD plasma phase close to $T_c$ in 
terms remnants of the confinement part of the QCD force rather than a 
strong Coulombic force.
\end{abstract}
\maketitle

\section{Introduction}
On quite general grounds it is expected, that fundamental forces
between quarks and gluons get modified at finite temperature. More 
precisely, we expect that forces between static quarks, $i.e.$ static 
test charges in a thermal medium, change because the gluons, which mediate 
the interaction between the static quarks, also interact with the 
constituents (quarks and gluons) of the thermal bath. In particular,
above the deconfinement temperature, $T_c$, the potential is expected
to be exponentially screened at large distances ($r \gg 1/T$) \cite{Linde}. 
In leading order perturbation theory this happens due to the generation
of a chromoelectric (Debye) mass of order $gT$ (with $g$ being
the gauge coupling). Beyond leading order, however, 
chromoelectric and chromomagnetic screening effects cannot be 
separated unambiguously. It is this non-perturbative large distance 
physics, which plays a central role in our attempts to understand the bulk
properties of the QCD plasma phase, {\it e.g.} the equation of state and 
the apparent deviations from ideal gas behavior found in numerical
calculations \cite{Boyd}. 
On the other hand it is the short and intermediate
distance regime which is most important in the discussion of 
signals which are considered today as being suitable to gain 
information on properties of hot and dense matter generated
experimentally in heavy ion collisions. 
In this paper
we will quantify the temperature dependence of the length scale which
separates these different regimes and analyze in detail the properties
of the QCD coupling constant at short and large distances. 

Although a detailed understanding of screening phenomena at 
large distances is still missing, it is evident that in this regime 
the temperature is the dominant scale and consequently will control the 
running  of the QCD coupling, {\it i.e.} $g\simeq g(T)$ for 
($rT \gg 1$, $T\gg T_c$)\footnote{We use the deconfinement temperature 
$T_c$ as a characteristic energy scale rather than a more conventionally
used $\Lambda$-parameter.}. However, at short distances, 
$r\cdot \max (T,T_c)\ll 1$, hard processes dominate the physics 
of the quark gluon plasma even at high temperature and it is expected that 
a scale appropriate for this short distance regime will control
the running of the QCD coupling, {\it i.e.} $g\simeq g(r)$. 
The interplay between short and large distance length scales plays a 
crucial role for a quantitative understanding of hard as well as soft 
processes in dense matter. It will, for instance, determine the range
of applicability of perturbative calculations for thermal dilepton rates 
or the production of jets as well as  the analysis of processes that can 
lead to thermalization of the dense matter produced in heavy ion collisions. 
Moreover, the short and intermediate distance regime also is most
relevant for the discussion of in-medium modifications of heavy quark bound 
states which are sensitive to thermal modifications of the heavy quark
potential as well as the role of quasi-particle excitations in the 
quark-gluon plasma. In all these cases it is not immediately evident that 
temperature is the relevant scale that controls the running of the QCD 
coupling at energies currently relevant in heavy ion physics where
temperatures may be reached which are only moderately larger than the phase 
transition temperature $T_c$. An analysis of this question becomes of
particular interest in view of the recently suggested scenario for the
existence of a large number of Coulombic bound states in
the QCD plasma phase close to $T_c$ \cite{Shuryak,Brown}. 

It is the purpose of this paper, to firmly establish that also at finite
temperature  the QCD coupling indeed runs as function of the length scale $r$ 
and agrees with the zero temperature running coupling at sufficiently
short distances. 
In fact, we will show that in the entire regime of distances for which at
zero temperature the heavy quark potential is considered to be 
described well by QCD perturbation theory \cite{Peter,Schroder,Melles} 
($r\; \lsim\; 0.1$~fm) the QCD
coupling remains unaffected by temperature effects up to
$T\simeq 3 T_c$. Furthermore, we will analyze the interplay
between the two relevant scales controlling the behavior of heavy
quark free energies, $F(r,T)$, and
we will quantify the distance scale below which processes in 
the QCD plasma phase are still  
dominated by properties of the QCD vacuum and above which screening
dominates the physics in the plasma phase.  

We will start in section II with a discussion of 
heavy quark free energies and their relation to the QCD coupling constant
and give some details on our simulation parameters in section III.
In section IV we
discuss the calculation of running couplings at finite temperature 
and short distances from color singlet free energies  
and relate these results to properties of 
the running coupling at large distances in section V. Section VI
contains our conclusions. 

\section{Free energies and running couplings}

\subsection{Free energy of a static quark anti-quark pair in lattice QCD}

Our main concern in this study is the determination of a running coupling
from correlation functions of Polyakov loops. In particular we 
want to make contact to calculations performed at zero temperature. Here the
running of the QCD coupling has been determined successfully in lattice 
calculations from the short distance properties of the heavy quark potential
\cite{Bali,necco01,necco02} and contact could be made with perturbative 
results obtained in 2-loop calculations \cite{Peter,Schroder}.
We thus want to analyze a finite temperature observable which naturally is 
related to the zero temperature heavy quark potential and has a 
well defined (perturbative) interpretation that allows us to make contact 
with perturbative definitions of a running coupling. For this reason
we will analyze properties of heavy quark free energies in the singlet
channel. 

Correlation functions of Polyakov loops define the free energy of a heavy
quark anti-quark pair. One generally considers the so-called color averaged
free energy,
\be
e^{-F(r,T)/T+C}=\frac{1}{9}\langle \tr L(\bx) \tr L^{\dagger}(\b0) \rangle \; ,
\label{f}
\ee
where the Polyakov loop, $L(\bx)$, is defined on lattices with temporal
extent $N_\tau$ in terms of temporal link variables $U_0(\bx,\tau)\in SU(3)$,
\be
L(\bx)=\prod_{\tau=1}^{N_{\tau}} U_0(\bx,\tau)\quad .
\label{Polyakov}
\ee
Furthermore, $r=|\bx|$ and $C$ is a suitably chosen renormalization constant, 
which can be determined from a matching of finite temperature free energies to
the zero temperature heavy quark potential \cite{okacz02,pisarski}.  

The color averaged free energy can be considered as a thermal average over 
contributions corresponding to quark anti-quark sources in color singlet and 
octet states, respectively, \cite{mclerran81,nadkarni86}
\be
e^{-F(r,T)/T}={1\over 9} e^{-F_1(r,T)/T}+{8\over 9} e^{-F_8(r,T)/T}\quad .
\label{f18}
\ee
where
\bea
e^{-F_1(r,T)/T+C}&=&{1\over 3} \tr \langle  L(\bx) L^{\dagger}(\b0) \rangle 
\quad ,
\label{f1}\\
e^{-F_8(r,T)/T+C}&=&{1\over 8}\langle \tr  L(\bx) \tr L^{\dagger}(\b0)\rangle- 
\nonumber \\
&& {1\over 24} \tr \langle L(\bx) L^{\dagger}(\b0) \rangle\quad .
\label{f8}
\eea
While the color averaged free energy is defined in terms of a 
gauge invariant Polyakov loop correlation function,
the singlet and octet correlation functions are given in terms 
of a gauge dependent correlator, $\displaystyle{\tr  L(\bx) 
L^{\dagger}(\b0) }$, and thus have to be evaluated in a fixed gauge. 
Nonetheless, $\displaystyle{\tr L(\bx) L^{\dagger}(\b0) }$ is
related to an appropriately chosen gauge invariant correlator and thus
has a proper gauge invariant interpretation. In fact, when evaluated in Coulomb
gauge the singlet correlation function constructed from the Polyakov loops 
defined in Eq.~\ref{Polyakov} may be viewed as resulting from gauge 
fixing non-local but gauge invariant operators, {\it i.e.} {\it dressed} 
Polyakov loops, where the static quark and anti-quark sources
are surrounded by gluon clouds \cite{vink,ophil02}.  

For our purpose of defining a running coupling at finite temperature the
singlet free energy is most appropriate as it has at short ($r\cdot \max
(T,T_c) \ll 1$) as well as large  $(rT\gg 1, T \gg T_c)$ distances and
temperatures a simple asymptotic behavior which is dominated by
one gluon exchange\footnote{Although this is usually called the
leading order perturbative result in the high temperature phase
the screened potential already involves summation of an infinite set of 
ladder diagrams which leads to the screening mass $\mu(T) = g(T) T$. We 
note, that this leading order perturbative result is gauge invariant.}, 
{\it i.e.}
\begin{equation}
\hspace*{-0.1cm}F_1(r,T) =
\cases{
-\;\frac{g^2(r)}{3\pi r} &\hspace*{-0.2cm}, $r \cdot \max (T,T_c) \ll 1$ \cr
~&~\cr
-\; \frac{g^2(T)}{3\pi r}{\rm e}^{-g(T)r T} &\hspace*{-0.2cm}, $(rT \gg 1, T \gg T_c)$}
\label{f1_asymp}
\end{equation}
Here we have already anticipated the running of the couplings with the
dominant scales in both limiting regimes, although their running,
of course, only arises in higher order perturbative
calculations. We also have suppressed any additative constants which
result from the renormalization of the free energy  \cite{okacz02} and, 
in particular, at high temperature will dominate the free energy in 
the large distance limit.  

While the relation between the singlet free energy and a running coupling
is straightforward the definition of a running coupling with the help of the
color averaged free energy is problematic.
The exact cancellation of leading order perturbative terms in high
temperature perturbation theory, $F(r,T)/T \sim (F_1(r,T)/T)^2$, which
generically does not occur at finite distances and temperatures,
makes it difficult to define a running coupling which easily could be
motivated by perturbation theory. While the above quadratic relation
between $F(r,T)$ and $F_1(r,T)$
holds at large distances it has been demonstrated by us recently that even
at high temperature  the short distance part of the color averaged
free energy is dominated by the singlet contribution and 
$F(r,T) \sim F_1(r,T)$ holds at short distances \cite{okacz02}.
In fact, for this reason the determination of a  
screening mass from the exponential fall of $F(r,T)$ at 
large distances, {\it i.e.} $F(r,T) \sim \exp \{- \mu (T) r \}$, 
turned out to be quite difficult and strongly dependent on the 
asymptotic form used in fits of the large distance behavior of $F(r,T)$
\cite{Kaczmarek}.
The sub-leading power-like corrections were found to be 
strongly temperature dependent and turned out to be difficult to control.

\subsection{Running couplings}

The perturbative short and large distance relations for the singlet 
free energy will be used to define a running coupling at finite temperature. 
In general, the definition of a running coupling in QCD is not unique
beyond the validity range of 2-loop perturbation theory; aside from the
scheme dependence of higher order coefficients in the QCD
$\beta$-functions it will strongly depend on non-perturbative
contributions to the observable used for its definition. This is quite
apparent when defining the coupling in QCD either in terms
of the free energy ($T=0$: potential)
\be 
\alpha_{\rm V} (r,T) = -\frac{3r}{4} F_1(r,T) \quad ,
\label{alpha_V}
\ee 
or its derivative ($T=0$: force)
\begin{equation}
\alpha_{\rm qq} (r,T) = \frac{3r^2}{4} \frac{{\rm d} F_1(r,T)}{{\rm d}r}
\quad .
\label{alpha}
\end{equation}
At low temperature the former necessarily has to change sign at
some intermediate distance due to the dominance of the linearly rising
confinement part in the potential \cite{Grunberg}. The latter, however, 
stays positive as $F_1(r,T)$ increases monotonically with $r$. In fact,
for a linear confining potential $\alpha_{\rm V}$ defined through
Eq.~\ref{alpha_V} will become negative
and drop quadratically while $\alpha_{\rm qq}$ defined through
Eq.~\ref{alpha} will rise
quadratically at large distances. The latter gives the possibility of 
smoothly matching 
the increasing coupling in the perturbative regime to the non-perturbative 
increase. This and the poor convergence of the perturbative expansion
for $\alpha_{\rm V}(r) $ have been reasons for analyzing in lattice 
calculations running couplings defined through Eq.~\ref{alpha}
($qq$-scheme) rather than Eq.~\ref{alpha_V} ($V$-scheme). 
We will in the following consider both definitions as this will also 
help to distinguish the short and large distance regimes at finite
temperature.

In the
perturbative regime different definitions of the running coupling
are uniquely related through the QCD $\beta$-function,
\begin{eqnarray}
\alpha_{\rm qq}(r,T) &=& \alpha_{\rm V} (r,T) - 
r \frac{{\rm d}\alpha_{\rm V} (r,T)}{{\rm d} r} \nonumber \\
&\equiv& \alpha_{\rm V} (r,T) + \frac{g(r,T)}{2\pi} \beta (g)\quad ,
\label{beta}
\end{eqnarray}
with $\beta (g) = -b_0 g^3 -b_1 g^5- b_2 g^7 + {\cal O}(g^9)$ and 
universal coefficients $b_0=11/16\pi^2$, $b_1=102/(16\pi^2)^2$.
In higher orders the coefficients of the $\beta$-function are
scheme dependent. At zero temperature the 
heavy quark potential has been calculated in 2-loop perturbation theory
\cite{Peter,Schroder,Melles}. From these calculations the 3-loop coefficient, 
$b_2$, in the $V$-scheme \cite{Schroder} and the $qq$-scheme
\cite{necco01} could be extracted and allowed for a detailed comparison 
of running couplings determined in perturbative and non-perturbative lattice 
calculations \cite{necco01,necco02}. Good agreement
has been found at distances $r\; \lsim\; 0.1 {\rm fm}$, which also
has been estimated to be the range of validity of the perturbative 
calculations. At larger distances, however, any perturbatively motivated
definition of the running coupling will also become sensitive to
non-perturbative effects and may lead to quite different results. 

We will extent here the zero temperature studies of the heavy quark
potential and the force between static charges to finite temperature. In
this case the appropriate observable is the heavy quark free energy and
its derivative.
At short distances we follow the approach used also at $T=0$ 
and introduce a running coupling by analyzing the $r$-dependence of 
the force between static quark anti-quark sources, Eq.~\ref{alpha}, 
and will compare it with the definition of a coupling in terms of
the potential, Eq.~\ref{alpha_V}.
At large distances we will determine a $T$-dependent running
coupling directly from fits of $F_1(r,T)$ which are motivated by the 
perturbative large distance form given
in Eq.~\ref{f1_asymp}. We will be more specific on this in Section~V.

\section{Simulation parameters}

We will analyze in the following properties of heavy quark anti-quark
pairs in a thermal heat bath of gluons, {\it i.e.}
we consider correlation functions of Polyakov loops in the SU(3) gauge 
theory (quenched QCD) at finite temperature calculated on
Euclidean lattice of size $N_\sigma^3 \times N_\tau$. All our 
calculations have been performed on lattices with spatial extent $N_\sigma = 32$
and $N_\tau = 4$, 8 and 16 using the tree level Symanzik-improved gauge 
action \cite{weisz}. It has been verified by us earlier \cite{okacz02}
that this choice of action and lattice parameters is sufficient to 
suppress finite volume effects and finite
size effects such as the breaking of rotational symmetry
in the analysis of correlation functions at short and intermediate distances.  
Our simulations have been performed in the
temperature range $T_c < T \le 12\; T_c$. The simulation parameters
are summarized in Table~\ref{tab:parameter}. The temperature scale for the 
Symanzik-improved action was obtained earlier from calculations of the 
string tension 
at zero temperature and a determination of the critical coupling for
the deconfinement transition on lattices with temporal extent $N_\tau =4$
and 6 \cite{beinlich99}. 

\begin{table}
\begin{tabular}{c|cccccc}
\hline
\hline
~ &  \multicolumn{3}{c}{$\beta$} \\
\hline
$T/T_c$ & $N_{\tau}=4$ & $N_{\tau}=8$ & $N_{\tau}=16$ \\
\hline
1.03(1) & 4.090 & - & - \\
1.05(2) & 4.100 & 4.5592 & - \\
1.10(1) & 4.127 &   -    & - \\
1.15(1) & 4.154 &   -    & - \\
1.20(3) & 4.179 & 4.6605 & - \\
1.24(1) & 4.200 &   -    & - \\
1.30(1) & 4.229 & 4.7246 & - \\
1.50(3) & 4.321 & 4.8393 & 5.4261 \\
1.60(2) & 4.365 & 4.8921 & - \\
1.68(2) & 4.400 &   -    & - \\
2.21(5) & 4.600 &   -    & - \\
3.0(1) & 4.839 & 5.4261 & - \\
6.0(3) & - & 6.0434 & - \\
9.0(3) & - & 6.3910 &  -  \\ 
12.0(5) & - & 6.6450 &  - \\ 
\hline
\hline
\end{tabular}
\caption{Parameters of the simulations on $32^3\times N_\tau$ lattices using 
the tree-level Symanzik-improved action.}
\label{tab:parameter}
\end{table}

The calculation of singlet free energies has been performed in Coulomb
gauge, which on the lattice is realized by maximizing 
$\displaystyle{\tr \sum_{\mu=1}^3 U_\mu(\bx,\tau}) $ in each time slice,
{\it i.e.} for fixed $\tau$. A residual gauge degree of freedom is 
fixed by demanding $\displaystyle{\sum_\bx U_0(\bx,\tau)}$ to be
independent of $\tau$. Typically we have analyzed the correlation
functions on 100-500 independent gauge field configurations.

As we are interested in the short distance behavior of the heavy quark 
free energy it is important to correct for the violation of rotational
symmetry which is most pronounced in this region. 
Following \cite{necco02} we have replaced $F(r,T)$ by $F(r_I,T)$ (similarly 
for $F_1$ and $F_8$) where $r_I$ relates the separation between the static
quark and anti-quark sources to the Fourier transform of the tree-level
lattice gluon propagator, $D_{\mu \nu}$, {\it i.e.}  
\be
r_I^{-1}=4 \pi \int_{-\pi}^{\pi}{d^3 k\over {(2 \pi)}^3}
\exp(i \vec{k}\cdot \vec{r}) D_{00}(k) \quad .
\ee
For the Symanzik-improved action the time-like component of 
$D_{\mu \nu}$ is given by 
\be
D^{-1}_{00}(k)=
4 \sum_{i=1}^3 \left( \sin^2\frac{k_i}{2}+\frac{1}{3} \sin^4\frac{k_i}{2}
\right) \quad .
\ee
This procedure removes most of the lattice artifacts as is evident from
the smooth short distance behavior of the free energies shown in
Fig.~\ref{fig:overall}.
Therefore in what follows we will always show free energies plotted 
versus $r_I$, but will suppress the subscript $I$.

\section{Running coupling constant at short distances}

\begin{figure}
\epsfxsize=87mm
\centerline{\epsffile{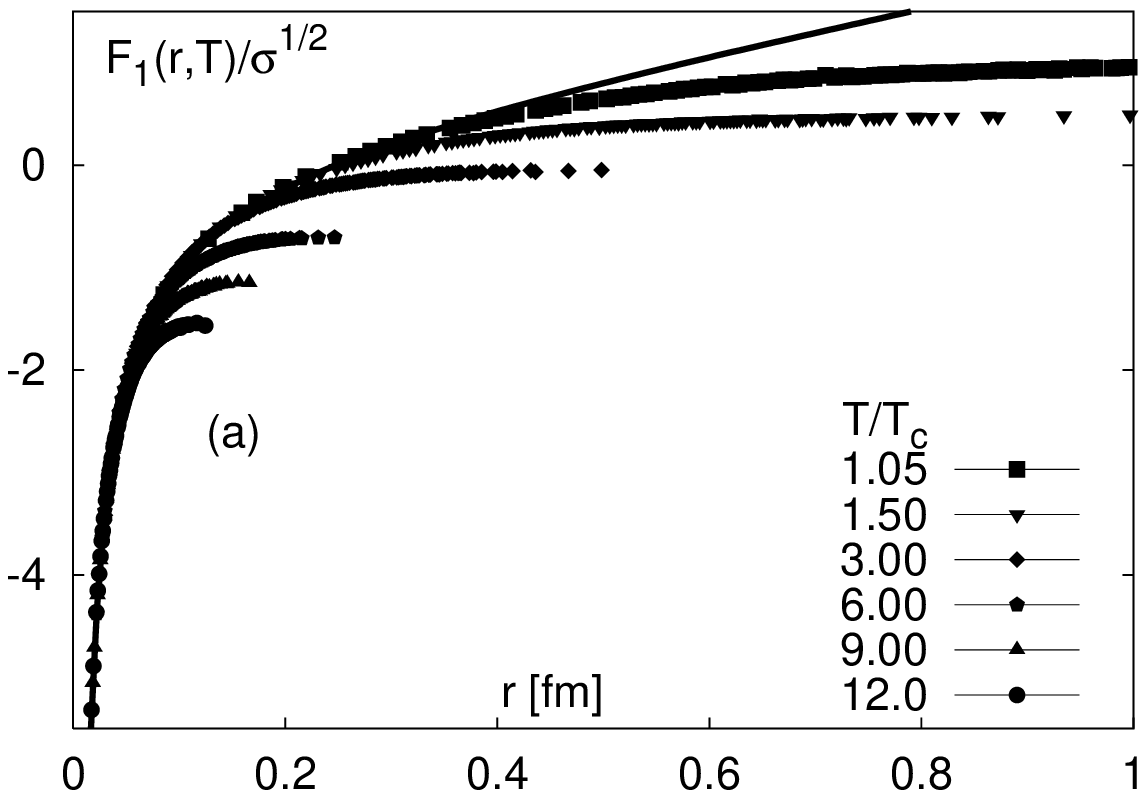}} 
\epsfxsize=90mm
\centerline{\epsffile{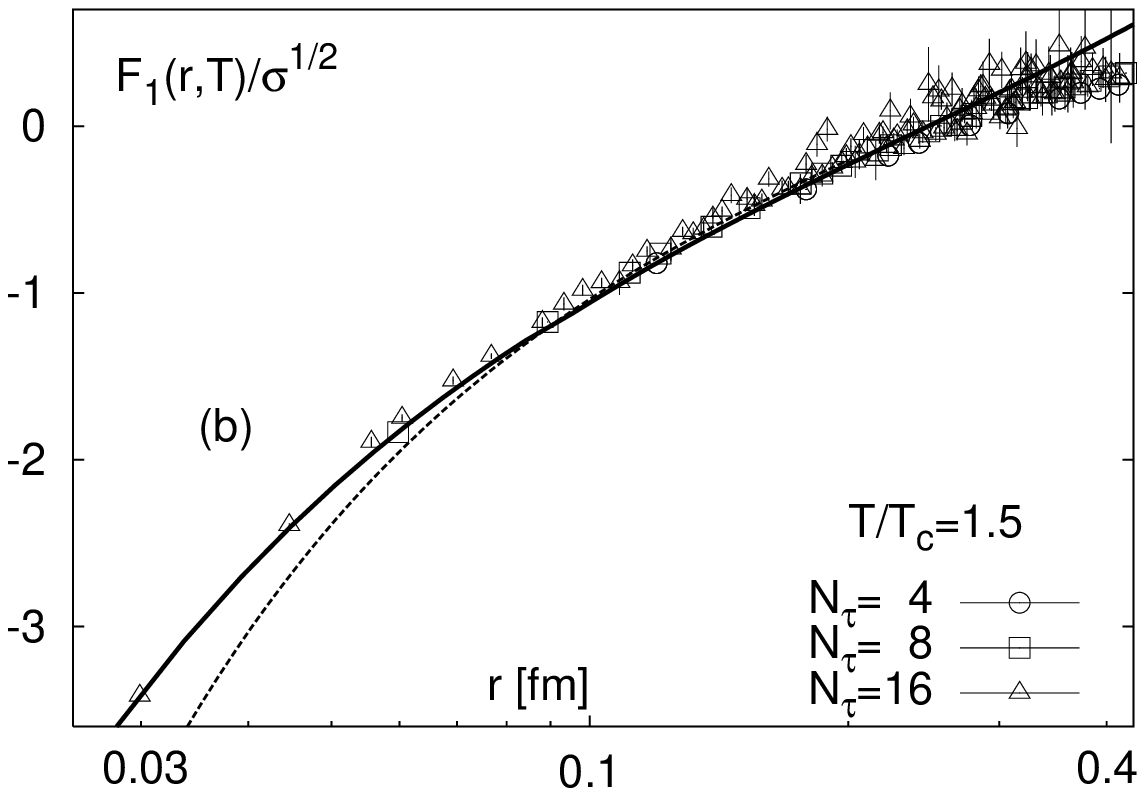}} 
\caption{Heavy quark free energies in the singlet  
channel calculated for several values of the temperature above $T_c$ (a) 
and the short distance part of the singlet free energy  at $T/T_c = 1.5$
(b). The dashed line shows $V_{string}(r)$ defined in
Eq.~\ref{Vstring} and the solid line is the perturbative result 
of Ref.~\cite{necco02} for $V_{\rm q\bar{q}}(r)$. This has been matched
smoothly to $V_{string}(r)$ at $r\simeq 0.1$~fm.
Physical units have been obtained by using $\sqrt\sigma = 420$~MeV.  
}
\label{fig:overall}
\end{figure}

In Fig.~\ref{fig:overall} we show our results for the 
singlet free energies for several temperatures above $T_c$. 
The free energies have been calculated in
Coulomb gauge and have been renormalized by matching 
the short distance part to the zero temperature heavy quark potential
of Ref.~\cite{necco02}. 

Fig.~\ref{fig:overall}b shows the short distance part of the singlet
free energy calculated at $T = 1.5 T_c$ on lattices with temporal extent
$N_\tau = 4$, 8 and 16. This corresponds to lattice spacings ranging
from $a \simeq 0.12$~fm down to $a \simeq 0.03$~fm. Apparently the 
short distance part of the singlet free energy agrees quite well with the 
zero temperature heavy quark potential including a perturbatively
calculated Coulomb term \cite{necco01},
\be
V_{\rm q\bar{q}}(r) = - \frac{4}{3} \frac{\alpha_{\rm V}(r)}{r}\quad ,
\label{Vqq}
\ee
and shows no significant cut-off
dependence. On the other hand, deviations from a confinement potential
with a constant Coulomb like term that arises from universal string 
fluctuations,
\be
V_{string}(r) = -\frac{\pi}{12\; r} \; + \sigma r \quad ,
\label{Vstring}
\ee
are clearly visible at these short distances. This already indicates
that the short distance behavior of the singlet free energy is consistent 
with a running coupling that is controlled by the quark anti-quark separation, 
$r$, and shows no or only little temperature dependence. We also note that
after having renormalized the free energy through a matching at short
distances the large distance behavior is completely fixed. Above $T_c$
the singlet free energy approaches a temperature dependent constant value 
at large distances which changes sign for $T\simeq 3\; T_c$. 

\begin{figure}
\epsfxsize=9cm
\centerline{\epsffile{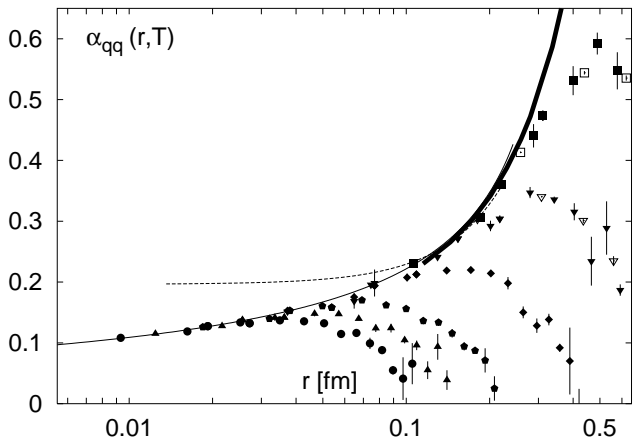}}
\caption{The running coupling in the $qq$-scheme determined on lattices of size
$32^3\times N_\tau$ with $N_\tau = 4$ (open symbols) and 8 (filled symbols)
from derivatives the short distance part of the singlet free energy
($T=0$: from the force)
at different temperatures. The relation of different symbols to the
values of the temperature are as in Fig.~1a. The various lines are explained 
in the text.
}
\label{alphaeff}
\end{figure}

To analyze the $T$ and $r$-dependence of the coupling we first 
follow the approach used at $T=0$ \cite{Bali,necco01} and define
$\alpha_{\rm qq} (r,T)$ through Eq.~\ref{alpha}. 
The derivatives of the singlet free energy with respect to the 
distance, ${\rm d} F_1(r,T)/{\rm d} r$, are obtained from a finite 
difference approximation using results at neighboring distances. 
We compare our finite temperature results to the high statistics
calculation performed at zero temperature \cite{necco02} 
in Fig. \ref{alphaeff}. In this figure the results obtained from the
numerical calculation of the heavy quark potential at $T=0$ and
distances $r\; \gsim\; 0.1$~fm are summarized by a fat black line. At
shorter distances a thin line represents the result of a
perturbative calculation of the force \cite{necco01,necco02} which is
based on the 2-loop calculation of the heavy quark potential
\cite{Peter,Schroder}. This perturbative result is smoothly matched
to the lattice data at $r\simeq 0.1$~fm. Also shown in the figure as a
dashed line is the effective coupling extracted from the confinement 
potential $V_{string}$ using Eq.~\ref{alpha}. 
It agrees quite well with the lattice data for $r\; \gsim\; 0.1$~fm but 
shows strong deviations from the perturbative as well as lattice calculation 
at shorter distances. 

Our numerical results on $\alpha_{\rm qq}$ at distances smaller than
$0.1$~fm cover also distances substantially smaller than those analyzed 
so far at $T=0$. They clearly show the running of the coupling with the 
dominant length scale $r$ also in the QCD plasma phase. For temperatures 
below $3 T_c$ we find that $\alpha_{\rm qq}$ agrees with the
zero temperature perturbative result in its entire regime of validity,
{\it i.e.} for $r\; \lsim\; 0.1$~fm. At these temperatures
thermal effects only become visible at larger distances 
and lead, as expected, to a decrease of the coupling
relative to its zero temperature value; for larger temperatures
thermal effects influence also the short distance behavior at 
distances $r\; \lsim\; 0.1$~fm. 

\begin{figure}
\epsfxsize=9cm
\centerline{\epsffile{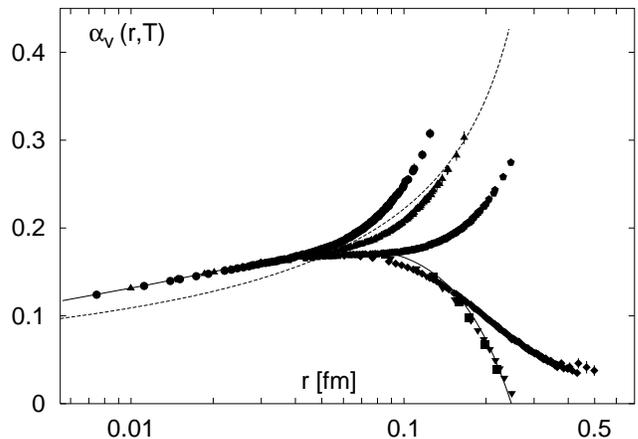}}
\caption{The running coupling in the $V$-scheme determined on lattices of size
$32^3\times N_\tau$ with $N_\tau = 4$ (open symbols) and 8 (filled
symbols)
from the short distance part of the singlet free energy ($T=0$:
from the potential)
at different temperatures. The relation of different symbols to the
values of the temperature are as in Fig.~1a. The dashed (solid) lines 
for $\alpha_{\rm qq}$ ($\alpha_{\rm V}$)
represent the zero temperature results of Ref.~\cite{necco02} which are
continued to shorter distances using our new results.
}
\label{fig:alpha_V}
\end{figure}

At distances larger than $r\; \simeq\; 0.1$~fm non-perturbative
effects clearly dominate the properties of $\alpha_{\rm qq}$. It thus is
to be expected that the properties of a running coupling will
strongly depend on the physical observable used to define it.
To quantify this we analyze directly the short distance behavior of the
renormalized singlet free energies and define $\alpha_{\rm V} (r,T)$
through Eq.~\ref{alpha_V}. At $T=0$ the singlet free energy simply is 
the heavy quark potential which for distances $r\; \gsim\; 0.1$~fm 
is quite well described by $V_{string}(r)$. 
In the case of the string potential, $V_{string}$, one would obtain
with the perturbative ansatz
given in Eq.~\ref{alpha_V} for the running coupling 
$\alpha_{\rm V}(r, 0) = \pi /16 - 0.75 \sigma r^2$, which 
changes sign at $r\simeq 0.25$~fm. We expect to find a similar behavior
also when using $F_1(r,T)$ at temperatures close to $T_c$.
As can be seen from Fig.~\ref{fig:alpha_V} such a 
behavior is indeed found for $T_c \le T \; \lsim\; 3T_c$. This 
again reflects the importance of remnants of the confining force in
the QCD plasma phase. For larger temperatures $\alpha_{\rm V}$ stays
positive reflecting the fact that $F_1(r,T)$ approaches a  negative
constant at large distances.   
We note that at short distance $\alpha_{\rm V}(r, T) > \alpha_{\rm qq} (r,T)$ 
as expected from the perturbative relation, Eq.~\ref{beta}, between
both definitions of the running coupling. At distances $r \; \gsim\; 0.1$~fm
the couplings, however, merely reflect the non-perturbative properties
of the observable used to define them.

\begin{figure}
\epsfxsize=9cm
\centerline{\epsffile{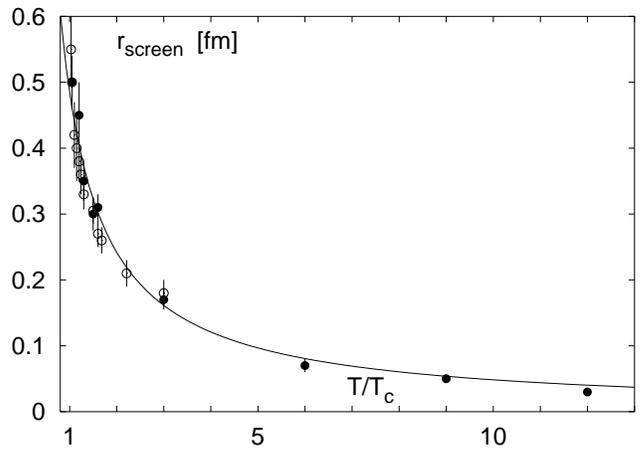}}
\caption{The location of the maximum in $\alpha_{\rm qq}(r,T)$
at fixed temperature versus temperature in units of the deconfinement 
temperature. Open (filled) symbols correspond to calculations on lattices
with temporal extent $N_\tau = 4$ and 8, respectively.
The solid line corresponds to 
$r_{\rm screen} = 0.48{\rm fm}\;\cdot T_c/T $.
}
\label{fig:scale}
\end{figure}

Let us return to an analysis of the properties of the running coupling
defined in the $qq$-scheme.
At temperatures close to $T_c$ the coupling $\alpha_{\rm V}(r,T)$ 
stays close to the zero temperature value up to distances 
$r\simeq 0.3$~fm. At these distances the strong $r$-dependence of
$\alpha_{\rm qq}(r,T)$ mimics the linear rising part of the 
zero temperature confinement potential. Remnants of the confining
force thus survive the deconfinement transition and play an important
role in quark anti-quark interactions at intermediate distances
up to $r\simeq 0.3$~fm. At distances larger than $\sim 0.5$~fm these
are, however, rapidly screened also at temperatures close to $T_c$.  
We use the maxima in $\alpha_{\rm qq}(r,T)$ to define a length scale
$r_{\rm screen}$ which separates the short distance regime from the
large distance regime. This is shown in Fig.~\ref{fig:scale}. In the entire 
temperature interval analyzed by us we find that $r_{\rm screen}$ is 
inversely proportional to the temperature, 
{\it i.e.} we find $r_{\rm screen} = (0.48\pm 0.01){\rm fm}\cdot T_c/T$. 

The fact that $\alpha_{\rm qq}(r,T)$ can become large at some distance
also in the deconfined
phase of QCD has recently been exploited to discuss the scenario of a 
strongly interacting fluid of quasi-particles describing the 
thermodynamics above but close to $T_c$ \cite{Shuryak,Brown}. Our analysis 
of the running coupling shows that up to a certain distance scale confining
features of the heavy quark potential indeed survive in the plasma phase and 
thus may support such a scenario. We note, however, that this effect is not 
related to an unexpectedly large coupling arising from thermal
effects but on the contrary to the survival of vacuum physics below
a certain characteristic length scale. In particular, there is no
evidence for a larger coupling in the Coulomb part of the heavy quark
potential arising from thermal effects. From this point of view it may
be questionable whether the plasma phase really can support the
existence of well localized colored bound states as it is advocated in 
\cite{Shuryak}.  

\section{$T$ dependence of the coupling at large distances}

\begin{figure}
\epsfxsize=9cm
\centerline{\epsffile{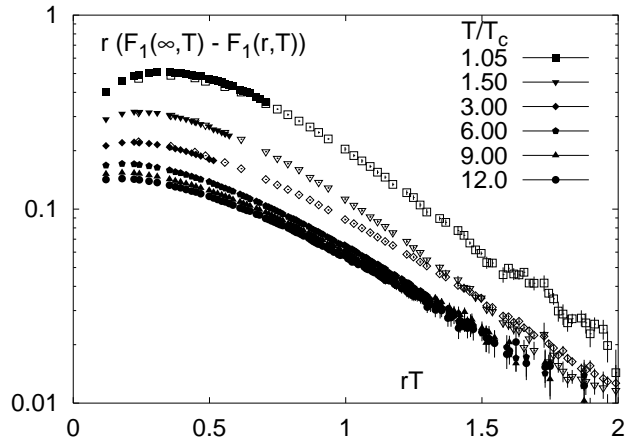}}
\caption{The singlet free energy versus $rT$ for several
temperatures above $T_c$ obtained from calculations on 
lattices
with temporal extent $N_\tau =4$ (open symbols) and $N_\tau =8$
(filled symbols).  
}
\label{fig:f1_log}
\end{figure}

We now turn to an analysis of the large distance structure of heavy
quark free energies and discuss the determination of a temperature 
dependent coupling from it. 
As inferred from the $r$-dependence of the coupling the crossover
from the short to large distance regime sets in rather
abruptly. In particular, the screening of the heavy quark free energies leads
to an exponential suppression of $\alpha_{\rm qq}$.
In order to extract the $T$-dependence of the 
QCD coupling conventionally used to describe the large distance
properties of QCD at high temperature this screening effect should 
be eliminated. In Fig.~\ref{fig:f1_log} we 
show $r (F_1(\infty,T)-F_1(r, T))$ on a logarithmic scale. 
Aside from deviations at
short and intermediate distances this is seen to decay exponentially. 
We fit the large
distance part of $F_1(r,T)$ with an ansatz motivated by the 
Debye screened perturbative result,
\be
\frac{F_{\rm fit} (r,T)}{T} = \frac{4\alpha(T)}{3 rT}\exp \{-\sqrt{4\pi
\tilde{\alpha}(T)}rT\}
+c(T)\quad ,
\label{fit}
\ee
where $\alpha(T)$ and $\tilde{\alpha}(T)$ are used as two independent fit 
parameters, which at large temperature, {\it i.e.} in the perturbative limit, 
are expected to coincide,
\begin{equation}
\lim_{T\rightarrow \infty} \alpha(T)/ \tilde{\alpha}(T) = 1\quad .
\end{equation}
We note that the coupling $\alpha(T)$ introduced here is 
closely related to the definition of $\alpha_{\rm V}$ given in
Eq.~\ref{alpha_V}. 

\begin{figure}
\epsfxsize=9cm
\centerline{\epsffile{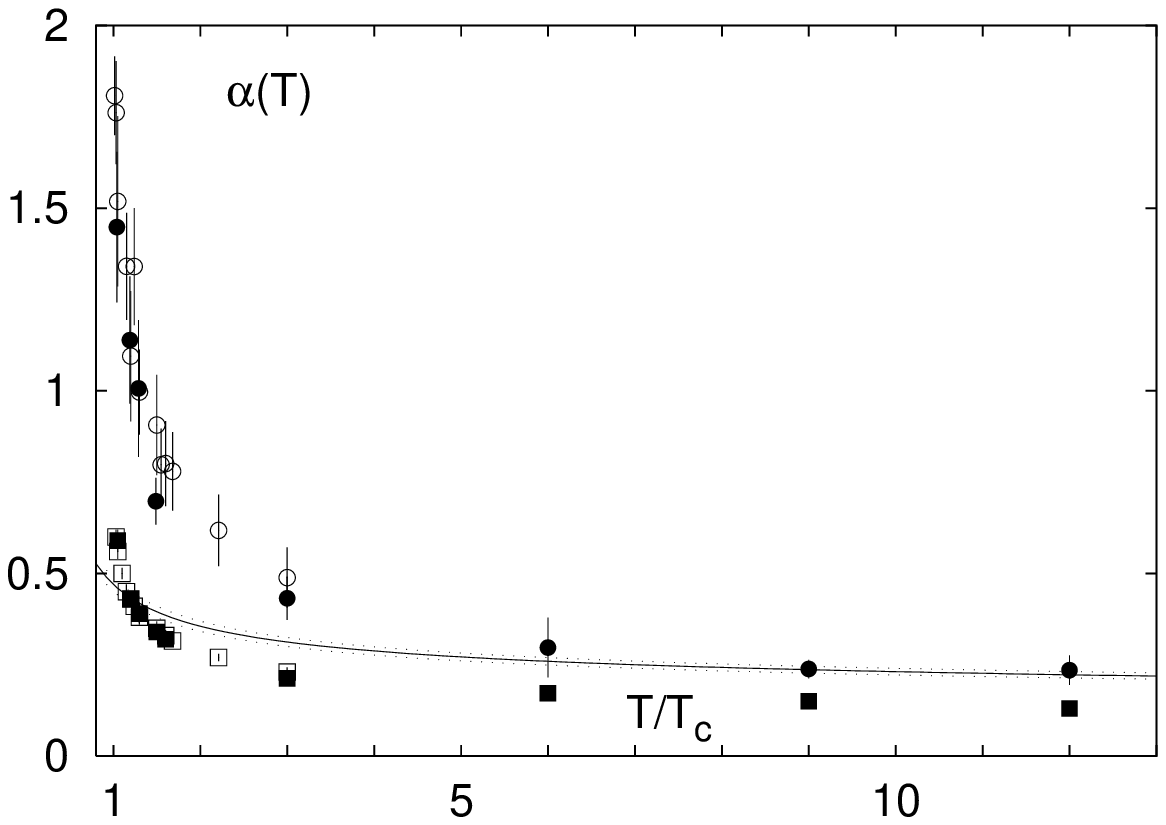}}

\epsfxsize=9cm
\centerline{\epsffile{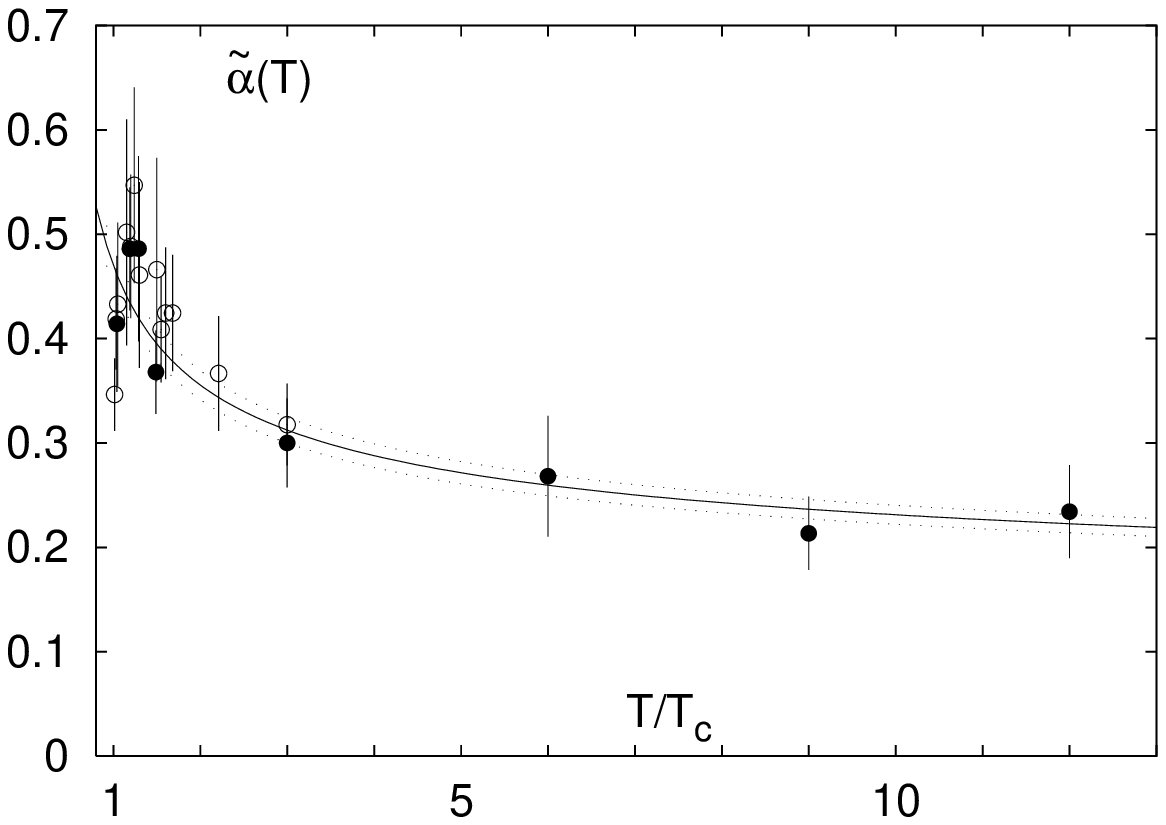}}

\epsfxsize=9cm
\centerline{\epsffile{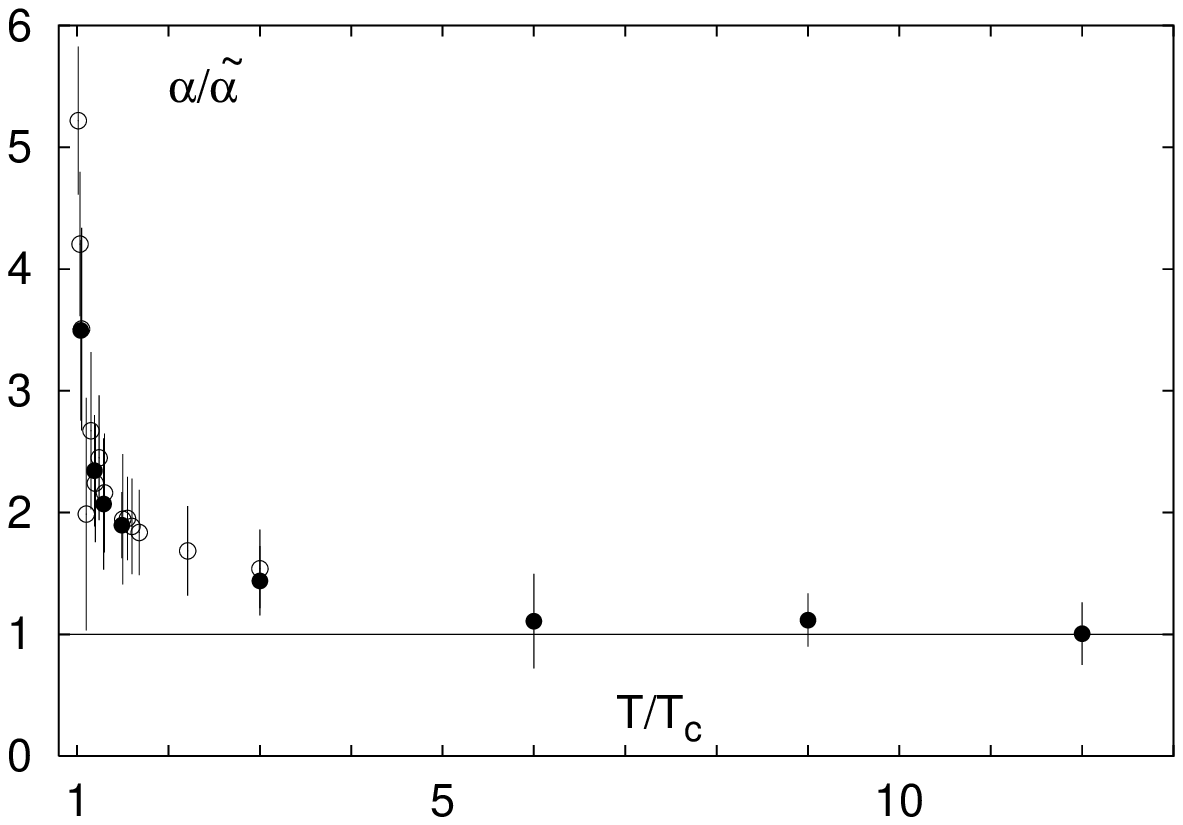}}
\caption{
The temperature dependent running coupling determined from the
large distance behavior of the singlet free energy on lattices
with temporal extent $N_\tau =4$ (open symbols) and $N_\tau =8$
(filled symbols). The upper 
figure shows $\alpha (T) \equiv g^2 (T)/4\pi$ (dots)
and the value $\alpha_{\rm qq} (r_{\rm screen}, T)$ (squares) determined from
the short distance behavior of the singlet free energy (see Fig.~3).
The figure in the middle shows $\tilde{\alpha} (T) \equiv \tilde{g}^2 (T)/4\pi$
and characterizes the temperature dependence of the screening mass.
The lower figure gives the ratio of both fit parameters.
The solid lines with the dotted error band are discussed in the text.
}
\label{fig:g}
\end{figure}

Both fit parameters as well as their ratio are shown in Fig.~\ref{fig:g}.
Within statistical errors the above limit 
indeed is reached for temperatures $T/T_c \simeq 6$. At smaller temperatures,
however, deviations from unity are large and reach 
$\alpha(T)/\tilde{\alpha}(T) \simeq 5$ close to $T_c$. 
Similar to the ambiguities that exist for the definition of a running
coupling at large distances this suggests that for not too large temperatures, 
$T_c\; \le\; T\; \lsim\; 2 T_c$, any definition of a 
temperature dependent running coupling 
will strongly depend on the physical process used to define $\alpha(T)$
or equivalently $g(T)$. 
Nonetheless for $T\; \gsim\; 6 T_c$ the temperature dependence seems to follow 
the logarithmic
behavior expected from perturbation theory. This is shown in the figure
by the solid lines with the dotted error band which represent an appropriately
rescaled perturbative running coupling, $\alpha (T) = 2.095 (82) 
\alpha_{\rm pert} (T)$ where we have used for $\alpha_{\rm pert}$ the
2-loop perturbative running coupling,
\be
g^{-2} (T) = {11 \over 8\pi^2} \ln \biggl({2\pi T\over
\Lambda_{\overline{\rm MS}}}\biggr) + {51 \over 88 \pi^2} \ln\biggl[
2\ln
\biggl( {2\pi T\over \Lambda_{\overline{\rm MS}}}\biggr) \biggr] \quad ,
\end{equation}
and relate $\Lambda_{\overline{\rm MS}}$ to the critical temperature
for the deconfinement transition,
$T_c/\Lambda_{\overline{\rm MS}} \simeq 1.14(4)$~\cite{Bali,Beinlich}. 
The rescaling factor has been determined from a common fit of the data
for $\alpha (T)$ and $\tilde{\alpha} (T)$ at temperatures $T\ge 6\; T_c$.
At least in this temperature regime all higher order perturbative as well
as non-perturbative 
effects seem to be well described by a rescaling of the coupling with
constant factor. A similar observation has been made for screening
masses determined in an SU(2) gauge theory at high temperature
\cite{Rank}.

Also shown in the upper part of Fig.~\ref{fig:g} is the maximal value of the
running coupling determined from the short distance behavior of the singlet 
free energy, $\alpha_{\rm qq} (r_{\rm screen}, T)$. As can be seen,
this coupling is significantly smaller than $\alpha (T)$ determined 
as coupling strength of the Debye screened Coulomb potential which
describes the long distance part of the singlet free energy. We stress
again that $\alpha (T)$ found here for $T\; \lsim\; 3T_c$ at large distances 
is not appropriate to characterize the Coulombic part of $F_1(r,T)$ at short 
distances. As discussed in the previous section this is still  
controlled by an almost temperature independent coupling $\alpha_V(r,T)$.

\section{Summary}
We have performed a detailed study of the singlet free energy
of a static $q \bar q$-pair in quenched QCD ($SU(3)$ gauge theory) 
at short and large distances. We have shown that at sufficiently
short distances the free energy agrees well with the zero 
temperature heavy quark potential and thus also leads to
a temperature independent running coupling. 
This short distance regime is temperature dependent 
and reduces from $r\simeq 0.5$~fm at $T\simeq T_c$ to
$r\simeq 0.03$~fm at $T\simeq 12\; T_c$. 
At high enough temperatures, $T\; \gsim\; 6T_c$,
the large distance behavior of the free energy is qualitatively similar 
to what is expected in perturbation theory
and allows for a consistent definition of a temperature dependent
running coupling characterizing large distance properties of
QCD thermodynamics. However, at temperatures close to $T_c$ the definition 
of a temperature dependent running coupling is not unique. 

Our analysis
suggests that it is more appropriate to characterize the non-perturbative
properties of the QCD plasma phase close to $T_c$ in terms remnants of
the confinement part of the QCD force rather than a strongly
coupled Coulombic force.

\section*{Acknowledgments}
This work has partly been supported by DFG under grant KA 1198/6-4
and GRK 881/1-04, by BMBF under grant No.06BI112 and partly by 
contract DE-AC02-98CH10886 with the U.S. Department of Energy.

\end{document}